# Surpassing the shot-noise limit by homodyne-mediated feedback

GUOFENG ZHANG(张国锋),* HANJIE ZHU(朱汉杰)

*Key Laboratory of Micro-Nano Measurement-Manipulation and Physics (Ministry of Education), School of Physics and Nuclear Energy Engineering, Beihang University, Xueyuan Road No. 37, Beijing 100191, China*
*Corresponding author: gf1978zhang@buaa.edu.cn*

**Entangled systems with large quantum Fisher information (QFI) can be used to outperform the standard quantum limit of the separable systems in quantum metrology. However, the interaction between the system and the environments inevitably leads to decoherence and decrease of the QFI, and it is not clear whether the entanglement systems can be better resource than separable systems in the realistic physical condition. In this work, we study the steady QFI of two driven and collectively damped qubits with homodyne-mediated feedback. We show that the steady QFI can be significantly enhanced both in the cases of symmetric feedback and nonsymmetric feedback, and the shot-noise limit of separable states can be surpassed in both cases. The QFI can even achieve the Heisenberg limit for appropriate feedback parameters and initial conditions in the case of symmetric feedback. We also show that an initial-condition-independent steady QFI can be obtained by using nonsymmetric feedback.**

An important application area that quantum systems can provide advantages is metrology, where some entangled multipartite quantum states can achieve better sensitivity in parameter estimation than separable states [1,2]. In quantum metrology, the precision of estimating a parameter is limited by the quantum Cramer-Rao inequality, and the quantum Fisher information (QFI) of the probe system imposes an upper bound on it [3,4]. A larger QFI means a higher precision on estimating the parameter. For a separable system of $N$ particles, the QFI can be at most $N$ which is referred as the shot-noise limit (SNL). The entangled system can surpass SNL and the QFI can achieve $N^2$, which is called the Heisenberg Limit (HL) [5]. However, the entanglement alone is not sufficient for surpassing SNL, i.e., be useful for parameter estimation. It has been found that not all entangled states can exceed SNL [6]. The QFI of a state can even increase as its entanglement decrease [7]. Due to the deep but not direct connection between QFI and entanglement, QFI has been studied for types of multiparticle entangled state [7-12]. Besides, the quantum system will inevitably interact with the surrounding environment, which leads to decoherence and decrease of the QFI [13-15]. In order to obtain a useful system under the realistic condition, various methods have been proposed to suppress the decoherence, such as quantum error correction, dynamical decoupling, reservoir engineering [16-23].

An alternative way to suppress the decoherence is quantum feedback, which manipulates the system based on the measurement results. For example, the entanglement of the steady state can be improved by applying a Markovian feedback [24]. Recently, a feedback scheme based on photodetection measurement has been proposed to enhance the QFI in the system that the qubit couples to the cavity [25]. Therefore, it is natural to ask whether the quantum feedback can excite QFI as it excites entanglement. In this work, we explore the possibility of enhancing the QFI and making the system useful by quantum feedback in the two qubits case. We will show that the quantum feedback can greatly enhance the QFI and help the system to be a better resource in quantum metrology.

First we would like to give a brief introduction about the QFI. In order to estimate a parameter $\phi$, one usually prepares a probe system, exposes it to the field associated with the parameter $\phi$ for a certain time, and measures the probe. The parameter $\phi$ can be estimated by comparing the input and output of the probe. In a linear interferometer, the probe system is experienced a phase transformation [26,27]

$$\rho_{out} = e^{-i\phi J_{\vec{n}}} \rho e^{i\phi J_{\vec{n}}}, \quad (1)$$

where $\rho$ and $\rho_{out}$ is the input and output state of the probe system respectively, $\phi$ is the phase shift, and $J_{\vec{n}}$ is the component of the collective angular momentum in the direction $\vec{n}$. The phase shift $\phi$ is estimated from the measurement results of $\rho_{out}$. The phase estimation sensitivity is limited by the quantum Cramer-Rao bound as

$$\Delta\hat{\phi} \geq \frac{1}{\sqrt{N_m F(\rho, J_{\vec{n}})}}, \quad (2)$$

where $N_m$ is the number of experiments, $F(\rho, J_{\vec{n}})$ is the quantum Fisher information and the estimator $\hat{\phi}$ satisfying $\langle\hat{\phi}\rangle = \phi$. Then the maximal QFI per qubit over all directions of state $\rho$ is given as $F(\rho) = c_{max}/N$ [7]. Here $N$ is the number of qubits and $c_{max}$ is the largest eigenvalue of the symmetric matrix which elements are given as

$$C_{kl} = \sum_{i \neq j} \frac{(\lambda_i - \lambda_j)^2}{\lambda_i + \lambda_j} \left[ \langle i|J_k|j\rangle\langle j|J_l|i\rangle + \langle i|J_l|j\rangle\langle j|J_k|i\rangle \right], \quad (3)$$

where $\lambda_i$ and $|i\rangle$ are the eigenvalues and the corresponding eigenvectors of the density matrix of the state and $i \in \{x, y, z\}$. Here we have $F(\rho) > 1$ for the QFI surpasses SNL and $F(\rho) = N$ for QFI reaches HL.

In this study, we consider a probe system consisting of two identical qubits resonantly coupled to a single mode cavity which is driven and heavily damped. When the cavity mode can be adiabatically eliminated for the strong damping, the dynamical evolution of this system is described by a master equation followed by the density operator $\rho$, where $\rho$ only includes the two qubits [24,27]:

$$\frac{d\rho}{dt} = -i[H, \rho] + \mathcal{D}[A]\rho. \quad (4)$$

Here the cavity is driven by a laser field with Rabi frequency $\Omega$ and $H = \Omega(\sigma_{1x} + \sigma_{2x})$ represents the driving of the laser, where $\sigma_{i\alpha}$ is the Pauli matrix ($\alpha = x, y, z$) applied to the $i$th qubit ($i = 1, 2$). $\mathcal{D}$ is a superoperator defined as $\mathcal{D} = \mathcal{D}[A]\rho = A\rho A^+ - \{A^+A, \rho\}/2$ for irreversible evolution, and $A = -i\Gamma(\sigma_{1-} + \sigma_{2-})$ is the jump operator describing the interaction between the system and its environment, where $\sigma_- = \sigma_x - i\sigma_y$ is the lowering operator for the atom. $\Gamma = g^2/\kappa$ is the effective damping rate relevant to the atom-cavity interaction strength $g$ and cavity damping rate $\kappa$. For simplicity, we let $\Gamma = 1$.

Since the master equation (5) is symmetric with respect to the two atoms, the spaces spanned by the symmetric basis $\{|e\rangle = |ee\rangle, \ |s\rangle = (|ge\rangle + |eg\rangle)/\sqrt{2}, \ |g\rangle = |gg\rangle\}$ and antisymmetric basis $|a\rangle = (|ge\rangle - |eg\rangle)/\sqrt{2}$ are invariant subspaces. By letting $\dot{\rho} = 0$, we can obtain the steady solution of the master equation in the basis $\{|e\rangle, |s\rangle, |g\rangle, |a\rangle\}$ [28]

$$\rho_s = \begin{pmatrix} \frac{\Omega^4}{E_1} & -i\frac{2\sqrt{2}\Omega^3}{E_1} & -\frac{8\Omega^2}{E_1} & 0 \\ i\frac{2\sqrt{2}\Omega^3}{E_1} & \frac{\Omega^4 + 8\Omega^2}{E_1} & -i\frac{2\sqrt{2}\Omega(\Omega^2 + 8)}{E_1} & 0 \\ -\frac{8\Omega^2}{E_1} & i\frac{2\sqrt{2}\Omega(\Omega^2 + 8)}{E_1} & \frac{\Omega^4 + 8\Omega^2 + 64}{E_1} & 0 \\ 0 & 0 & 0 & \rho_{44}(0) \end{pmatrix}. \quad (5)$$

Here $E_1 = (3\Omega^4 + 16\Omega^2 + 64)/(1 - \rho_{44}(0))$, and $\rho_{ij}(0)$ ($i, j = 1,2,3,4$) are the elements of the initial density matrix. It is evident that the final steady state strongly depends on the probability $\rho_{44}(0)$ with which the initial state is in the antisymmetric subspace. If the probability $\rho_{44}(0)$ is given, the steady solution is independent of other initial condition of the system. In Fig. 1, we show the maximal QFI per particle as a function of the Rabi frequency $\Omega$ and probability $\rho_{44}(0)$. From Fig. 1 we find that if the initial state is restricted in the symmetric subspace, the QFI of the model can stay near the shot-noise limit and even slightly surpass this limit. For the optimal value of $\Omega$, the QFI per qubit can reach a maximum of 1.085. Besides, as the antisymmetric subspace probability $\rho_{44}(0)$ increased, the QFI of the model drops rapidly and eventually reaches zero if the initial state is in the antisymmetric subspace. It is a surprising result since the QFI exhibits the opposite behavior of the entanglement, while the entanglement of the steady state increases with the probability $\rho_{44}(0)$ [28]. Based on the above, we can draw a conclusion that without the feedback this model cannot be useful even in the optimal condition.

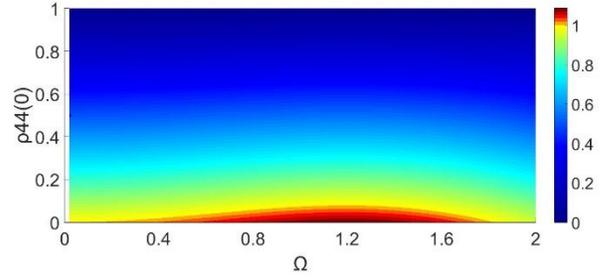

Fig. 1. The maximal quantum Fisher information per qubit as a function of the Rabi frequency $\Omega$ and probability $\rho_{44}(0)$ in the absence of feedback.

Now we consider subjecting the damping from the output of the cavity to a homodyne measurement with a Markovian feedback to the cavity. The feedback Hamiltonian can be written as $H_{fb} = I(t)F$, where $I(t)$ is the homodyne detection photocurrent. The feedback operator $F$ represents the additional controlled potential for the qubits. In this scheme the master equation is described by [28,29]

$$\frac{d\rho}{dt} = -i\left[H + \frac{1}{2}(A^+F + FA), \rho\right] + \mathcal{D}[A - iF]\rho. \quad (6)$$

With this master equation we can show how the quantum feedbacks enhance the QFI.

First we consider the symmetric feedback Hamiltonian $F_1 = \lambda[\mu(\sigma_{1x}\sigma_{2z} + \sigma_{1z}\sigma_{2x}) - (\sigma_{1x} + \sigma_{2x})]$. This corresponds to having a feedback-modulated driving laser when $\mu = 0$. It is easy to find that this feedback Hamiltonian preserves the symmetry with respect to the two qubits. Follow the same sort of procedure in the non-feedback case we could obtain the steady solution of the master equation. For simplicity, we choose $\Omega = 0$. Then the steady solution of the master equation in the basis $\{|e\rangle, |s\rangle, |g\rangle, |a\rangle\}$ takes the form [28]

$$\rho_{11} = \frac{(-1+\mu^2)^2 \lambda^4}{E_2},$$

$$\rho_{13} = -\frac{2(-1+\mu^2)(-1+\lambda)\lambda^2(1+\lambda\mu)}{E_2},$$

$$\rho_{22} = \frac{(1+\mu)^2 \lambda^2 \left[2 + 2(-1+\mu)\lambda + (1+\mu^2)\lambda^2\right]}{E_2},$$

$$\rho_{33} E_2 = 8 + 8(-3+\mu)\lambda + 8(3-2\mu+\mu^2)\lambda^2 \quad (7)$$
$$- 8(1-\mu+2\mu^2)\lambda^3 + (1+6\mu^2+\mu^4)\lambda^4,$$

$$\rho_{12} = \rho_{23} = \rho_{14} = \rho_{24} = \rho_{34} = 0, \rho_{44} = \rho_{44}(0),$$

where $E_2 = ((3 + 2\mu + 6\mu^2 + 2\mu^3 + \mu^4)\lambda^4 + 2(-5 + 3\mu - 7\mu^2 + \mu^3)\lambda^3 + 2(13 - 6\mu + 5\mu^2)\lambda^2 + 8(-3 + \mu)\lambda + 8)/(1 - \rho_{44}(0))$. This solution is valid for all values of $\mu$ and $\lambda$ except for $\mu = -1, \lambda = 1$ and $\mu = 1, \lambda = 1$. In Fig. 2 we show the maximal QFI per particle as a function of feedback parameters $\lambda$ and $\mu$ for the initial state restricted in the symmetric subspace. When $\mu = 0$, the model reduces to the feedback scheme [24]. In this case we can find that the QFI stays below the shot-noise limit, thus the system cannot be useful in this feedback scheme. In the $\mu \neq 0$ case, the QFI can be greatly enhanced for proper values of $\lambda$ and $\mu$. It can even achieve the Heisenberg limit when $\lambda = 1$ and $\mu = 1$. Therefore, in the presence of feedback $F_1$ the QFI are significantly enhanced and the model becomes useful when the parameters $\lambda$ and $\mu$ both approach 1.

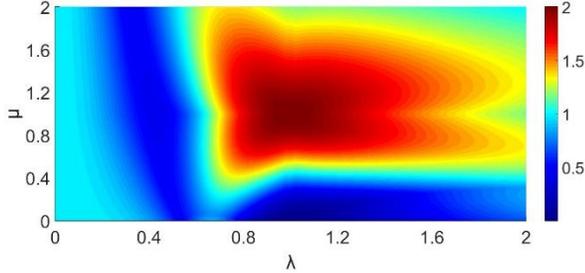

Fig. 2. The maximal quantum Fisher information per qubit as a function of feedback parameters $\lambda$ and $\mu$ in the case of symmetric feedback. Here we choose $\Omega = 0$.

Although the QFI can be excited in this symmetric feedback scheme, it strongly depends on the initial condition. If we allow the initial state to be in the full space, its antisymmetric subspace part will suppress the QFI. To illustrate this point we consider a specific case where $\lambda = 1$ and $\mu = 1$. In this case the steady solution (8) is valid except that the matrix element $\rho_{24}$ is no longer zero. It is easy to find that $\rho_{24}$ is time-independent and we have $\rho_{24} = \rho_{24}(0)$. Here we allow $\rho_{44}(0) \neq 0$ and choose $\rho_{24}(0) = 0$, then the steady state takes the form $\rho_s = (1 - \rho_{44}(0))|s\rangle\langle s| + \rho_{44}(0)|a\rangle\langle a|$. After some calculation, we obtain the maximal QFI of per particle as $F(\rho) = 2(1 - \rho_{44}(0))$, which is a decreasing function of $\rho_{44}(0)$. The QFI can reach the Heisenberg limit if the initial state is restricted in the symmetric subspace. Therefore, to obtain a steady state with high QFI, we need to prepare the initial state carefully and minimize the probability $\rho_{44}(0)$ with which the initial state is in the antisymmetric subspace.

In order to avoid the dependence of the QFI on the initial condition, we can introduce a nonsymmetric feedback which breaks the symmetry of the system. Here we choose the nonsymmetric feedback Hamiltonian as

$$F_2 = \lambda \left[ \mu(\sigma_{1x}\sigma_{2z} + \sigma_{1z}\sigma_{2x}) - ((1+\Delta)\sigma_{1z} + (1-\Delta)\sigma_{2z}) \right], \quad (8)$$

where there is a small difference $\Delta$ between the feedback amplitudes of two qubits. This small difference breaks the symmetry and leads to a steady solution which is independent of the initial condition. By choosing the basis $\{|ee\rangle, |eg\rangle, |ge\rangle, |gg\rangle\}$ and letting $\dot{\rho} = 0$, we can obtain 16 linear equations. Then the steady state can be obtained by solving these equation. Due to the complexity of the corresponding density matrix, we skip over the details of the expression and present the numerical results. In Fig. 3 we show the maximal QFI per particle as a function of feedback parameters $\lambda$ and $\mu$ for $\Delta = 0.1$. It can be seen that in the nonsymmetric case, the steady QFI can also be improved greatly for appropriate values of $\lambda$ and $\mu$. The QFI per qubit can reach a maximum value of 1.732 for $\lambda = 2$ and $\mu = 1$, which is far beyond the shot-noise limit. Therefore, we can find that no matter what initial condition the qubits are prepared in, the model can be useful by adjusting the feedback parameters.

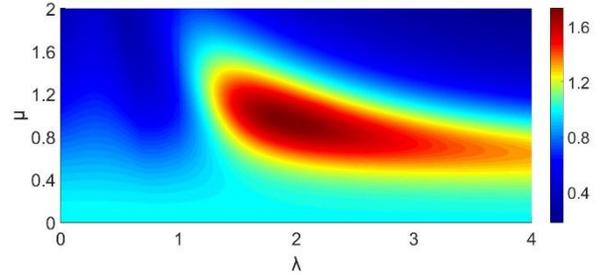

Fig. 3. The maximal quantum Fisher information per qubit as a function of feedback parameters $\lambda$ and $\mu$ in the case of nonsymmetric feedback. Here we choose $\Omega = 0$ and $\Delta = 0.1$.

To summarize, in the system of two collectively damped qubits, we demonstrate that the quantum feedback can enhance the quantum Fisher information and allow the system to be useful for surpassing the shot-noise limit in quantum metrology. We have shown that for the symmetric feedback case, the system can be useful by choosing appropriate feedback parameter. The QFI of the system can surpass the shot-noise limit and even reach the Heisenberg limit. However, in this feedback scheme the steady QFI is strongly depends on the initial condition, and we need to prepare the initial state carefully to minimize the probability with which the initial state is in the antisymmetric subspace.

We then present a nonsymmetric feedback scheme to overcome this initial-condition-dependent difficulty. In this case we can obtain the steady QFI which is independent of initial condition. We find that the QFI can also be excited in this scheme, and the system can become useful by adjusting the feedback amplitudes.

**Funding.** National Natural Science Foundation of China (Grant No. 11574022 and 11174024) and the Fundamental Research Funds for the Central Universities of Beihang University (Grant No. YWF-16-BJ-Y-29 and YWF-16-WLXY-001).

**REFERENCES**
1. C. M. Caves, Phys. Rev. D **23**, 1693 (1981).
2. V. Giovannetti, S. Lloyd, L. Maccone, Science **306**, 1330 (2004).

3. A. S. Holevo, *Probabilistic and Statistical Aspect of Quantum Theory* (North-Holland, 1982).
4. C. W. Helstrom, *Quantum Detection and Estimation Theory* (Academic Press, 1976).
5. L. Pezze, A. Smerzi, Phys. Rev. Lett. **102**, 100401 (2009).
6. P. Hyllus, W. Laskowski, R. Krischek, Phys. Rev. A **85**, 022321 (2012).
7. J. Ma, Y. X. Huang, X. Wang, C. P. Sun, Phys. Rev. A **84**, 022302 (2011).
8. M. Rosenkranz, D. Jaksch, Phys. Rev. A **79**, 022103 (2009).
9. F. Ozaydin, A. A. Altintas, C. Yesilyurt, S. Bugu, V. Erol, Act. Phys. Pol. A **127**(**4**), 1233 (2015).
10. W. F. Liu, J. Ma, X. Wang, J. Phys. A: Math. Gen. **46**, 045302 (2013).
11. X. M. Liu, Z. Z. Du, J. M. Liu, Quantum Inf. Process. **15**(**4**), 1793 (2016).
12. J. Ma, X. Wang, Phys. Rev. A **80**, 012318 (2009).
13. B. M. Escher, R. L. de Matos Filho, L. Davidovich, Nat. Phys. **7**, 406 (2011).
14. Y. M. Zhang, X. W. Li, W. Yang, G. R. Jin, Phys. Rev. A **88**, 043832 (2013).
15. F. Hudelist, J. Kong, C. Liu, J. Jing, Z. Y. Ou, W. Zhang, Nat. Commun. **5**, 3049 (2014).
16. U. Dorner, New J. Phys. **14**, 043011 (2012).
17. Y. Watanabe, T. Sagawa, M. Ueda, Phys. Rev. Lett. **104**, 020401 (2010).
18. Q. S. Tan, Y. X. Huang, X. L. Yin, L. Kuang, X. Wang, Phys. Rev. A **87**, 032102
19. A. W. Chin, S. F. Huelga, M. B. Plenio, Phys. Rev. Lett. **109**, 233601 (2012).
20. W. B. Dong, R. B. Wu, X. H. Yuan, C. W. Li, T. J. Tarn, Sci. Bull. **60**(**17**), 1493 (2015)
21. J. Jing, L. A. Wu, Sci. Bull. **60**(**3**), 328 (2015)
22. R. X. Xu, H. D. Zhang, X. Zheng, Y. J. Yan, Sci. China Chem. **58**(**12**), 1816 (2015)
23. R. J. Guo, J. S. Zheng, Y. S. Zhang, Y. C. Shi, L. Y. Li, B. C. Qiu, L. L. Lu, X. F. Chen, Sci. Bull. **60**(**11**), 1026 (2015)
24. J. Wang, H. M. Wiseman, G. J. Milburn, Phys. Rev. A **71**, 042309 (2005).
25. Q. Zheng, L. Ge, Y. Yao, Q. Zhi, Phys. Rev. A **91**, 033805 (2015).
26. P. Hyllus, O. Guhne, A. Smerzi, Phys. Rev. A **82**, 012337 (2010).
27. N. Yamamoto, Phys. Rev. A **72**, 024104 (2005).
28. J. G. Li, J. Zou, B. Shao, J. Cai, Phys. Rev. A **77**, 012339 (2008).
29. H. M. Wiseman, G. J. Milburn, Phys. Rev. A **47**, 642 (1993).

## References

1. C. M. Caves, *Quantum-mechanical noise in an interferometer*, Phys. Rev. D **23**, 1693 (1981).
2. V. Giovannetti, S. Lloyd, L. Maccone, *Quantum-enhanced measurements: beating the standard quantum limit*, Science **306**, 1330 (2004).
3. A. S. Holevo, *Probabilistic and Statistical Aspect of Quantum Theory* (North-Holland, 1982).
4. C. W. Helstrom, *Quantum Detection and Estimation Theory* (Academic Press, 1976).
5. L. Pezze, A. Smerzi, *Entanglement, Nonlinear Dynamics, and the Heisenberg Limit*, Phys. Rev. Lett. **102**, 100401 (2009).
6. P. Hyllus, W. Laskowski, R. Krischek, *Fisher information and multiparticle entanglement*, Phys. Rev. A **85**, 022321 (2012).
7. J. Ma, Y. X. Huang, X. Wang, C. P. Sun, *Quantum Fisher information of the Greenberger-Horne-Zeilinger state in decoherence channels*, Phys. Rev. A **84**, 022302 (2011).
8. M. Rosenkranz, D. Jaksch, *Parameter estimation with cluster states*, Phys. Rev. A **79**, 022103 (2009).
9. F. Ozaydin, A. A. Altintas, C. Yesilyurt, S. Bugu, V. Erol, *Quantum Fisher Information of Bipartitions of W States*, Act. Phys. Pol. A **127**(**4**), 1233 (2015).
10. W. F. Liu, J. Ma, X. Wang, *Quantum Fisher information and spin squeezing in the ground state of the XY model*, J. Phys. A: Math. Gen. **46**, 045302 (2013).
11. X. M. Liu, Z. Z. Du, J. M. Liu, *Quantum Fisher information for periodic and quasiperiodic anisotropic XY chains in a transverse field*, Quantum Inf. Process. **15**(**4**), 1793 (2016).
12. J. Ma, X. Wang, *Fisher information and spin squeezing in the Lipkin-Meshkov-Glick model*, Phys. Rev. A **80**, 012318 (2009).
13. B. M. Escher, R. L. de Matos Filho, L. Davidovich, *General framework for estimating the ultimate precision limit in noisy quantum-enhanced metrology*, Nat. Phys. **7**, 406 (2011).
14. Y. M. Zhang, X. W. Li, W. Yang, G. R. Jin, *Quantum Fisher information of entangled coherent states in the presence of photon loss*, Phys. Rev. A **88**, 043832 (2013).
15. F. Hudelist, J. Kong, C. Liu, J. Jing, Z. Y. Ou, W. Zhang, *Quantum metrology with parametric amplifier-based photon correlation interferometers*, Nat. Commun. **5**, 3049 (2014).
16. U. Dorner, *Quantum frequency estimation with trapped ions and atoms*, New J. Phys. **14**, 043011 (2012).
17. Y. Watanabe, T. Sagawa, M. Ueda, *Measuring entanglement in condensed matter systems*, Phys. Rev. Lett. **104**, 020401 (2010).
18. Q. S. Tan, Y. X. Huang, X. L. Yin, L. Kuang, X. Wang, *Enhancement of parameter-estimation precision in noisy systems by dynamical decoupling pulses*, Phys. Rev. A **87**, 032102
19. A. W. Chin, S. F. Huelga, M. B. Plenio, *Quantum Metrology in Non-Markovian Environments*, Phys. Rev. Lett. **109**, 233601 (2012).
20. W. B. Dong, R. B. Wu, X. H. Yuan, C. W. Li, T. J. Tarn, *The modelling of quantum control systems*, Sci. Bull. **60**(**17**), 1493 (2015)
21. J. Jing, L. A. Wu, *Overview of quantum memory protection and adiabaticity induction by fast signal control*, Sci. Bull. **60**(**3**), 328 (2015)
22. R. X. Xu, H. D. Zhang, X. Zheng, Y. J. Yan, *Dissipaton equation of motion for system-and-bath interference dynamics*, Sci. China Chem. **58**(**12**), 1816 (2015)
23. R. J. Guo, J. S. Zheng, Y. S. Zhang, Y. C. Shi, L. Y. Li, B. C. Qiu, L. L. Lu, X. F. Chen, *Suppressing longitudinal spatial hole burning with dual assisted phase shifts in pitch-modulated DFB lasers*, Sci. Bull. **60**(**11**), 1026 (2015)
24. J. Wang, H. M. Wiseman, G. J. Milburn, *Dynamical creation of entanglement by homodyne-mediated feedback*, Phys. Rev. A **71**, 042309 (2005).
25. Q. Zheng, L. Ge, Y. Yao, Q. Zhi, *Enhancing parameter precision of optimal quantum estimation by direct quantum feedback*, Phys. Rev. A **91**, 033805 (2015).
26. P. Hyllus, O. Guhne, A. Smerzi, *Not all pure entangled states are useful for sub-shot-noise interferometry*, Phys. Rev. A **82**, 012337 (2010).
27. N. Yamamoto, *Parametrization of the feedback Hamiltonian realizing a pure steady state*, Phys. Rev. A **72**, 024104 (2005).
28. J. G. Li, J. Zou, B. Shao, J. Cai, *Steady atomic entanglement with different quantum feedbacks*, Phys. Rev. A **77**, 012339 (2008).
29. H. M. Wiseman, G. J. Milburn, *Quantum theory of field-quadrature measurements*, Phys. Rev. A **47**, 642 (1993).